\documentclass[aps,prc,reprint,groupedaddress,showpacs,showkeys]{revtex4-1}

\usepackage{graphicx}
 \usepackage{epsfig}

\usepackage{amssymb}
\usepackage{amsmath}
\usepackage{lipsum}
\RequirePackage[colorlinks,citecolor=blue,urlcolor=blue,linkcolor=blue]{hyperref}
\usepackage{dcolumn}
\usepackage{footnote}
\begin{document} 


\preprint{PRC}

\title{Separate mass scaling of the widths of the rapidity distributions for mesons and baryons at energies available at the Facility for Antiproton and Ion research}



\author{Kalyan Dey}
\email[]{k.dey@gsi.de}
\author{B. Bhattacharjee}
\email[]{buddhadeb.bhattacharjee@cern.ch (Corresponding author)}

\affiliation{Nuclear and Radiation Physics Research Laboratory, Department of Physics, Gauhati University, Guwahati - 781014, India}


\date{\today}

\begin{abstract}
Evolution of the width of the rapidity distribution on beam rapidity has been studied for a number of produced particles with UrQMD-3.3p1 generated events at various FAIR (Facility for Antiproton and Ion Research) energies. The results for the width of the rapidity distribution with beam rapidity, obtained with UrQMD generated events, are compared with the existing experimental data (E802, E877, E896, E917, NA49). For both UrQMD and experimental data, the width of the rapidity distribution is found to bear scaling behavior with beam rapidity for all the hadrons. Such scaling behavior is found to follow separate mass ordering for the studied mesons and and baryons. 
\end{abstract}

\maketitle


\section{Introduction}
\label{A}

The upcoming Compressed Baryonic Matter (CBM) experiment at the future Facility for Antiproton and Ion Research (FAIR) will be a dedicated heavy ion experiment operating in fixed target mode ($Au+Au$ collision up to 35\textit{A} GeV)  and is planned to explore the properties of nuclear matter at moderate temperature and high baryon density 
\cite {peter,peter1,hone,volker}. In heavy ion collisions at FAIR energies, the baryon density is expected to be extremely high, something of the order of 5-10 times the normal nuclear matter density. As a result, effects which couple to baryon density are expected to be very prominent in this energy regime \cite{vogel}. One of the characteristic features of FAIR is its high luminosity beam, which would be suitable for probing such a hot and dense medium. The state-of-the-art large acceptance detectors of the CBM experiment will give it  access to almost the entire forward rapidity hemisphere. Thus, the evolution of the width of the rapidity distribution with beam energy and centrality will be experimentally addressed.

In this work, an attempt has been made, with UrQMD-3.3p1 (without taking into account the hydro part) \cite{urqmd1, urqmd2, petersen} generated $Au+Au$ events at 10, 20, 30, and 40\textit{A} GeV, to investigate the evolution of the width of the rapidity distribution with beam rapidity ($y_{b}$) of a few mesons and baryons and compare UrQMD predictions with the existing experimental data. The event statistics of our generated data (for central collisions only) are presented in Table \ref{event_statistics}. It may be worth mentioning that though there exist reports \cite{data5,phi_NA49_2} on the evolution of width of the rapidity distribution with beam rapidity inclusive of $\overline{\Lambda}$,  no such result has been reported including the $\Lambda$ baryon. Even though, $\Lambda$ and $\overline{\Lambda}$ have same mass, which is in close proximity with the mass of $\phi$ meson, a  distinctive feature of $\Lambda$ ($uds$) and $\overline{\Lambda}$ ($\overline{u}\overline{d}\overline{s}$) is that while the former is created from the leading baryons, the latter is created from newly produced quarks.

\begin{table*}[tb]
\setlength\extrarowheight{3pt}
\caption{Event statistics of the present investigation (for central collisions only).}
\label{event_statistics}
\begin{tabular*}{\textwidth}{@{\extracolsep{\fill}}ccrcrrrr@{}}
\hline
\hline
Energy & \multicolumn{1}{c}{Events} & \multicolumn{1}{c}{$\pi^{-}\times10^8$} & \multicolumn{1}{c}{$K^-\times10^7$} & \multicolumn{1}{c}{$\Lambda \times10^7$} & \multicolumn{1}{c}{$\phi \times10^5$} &  \multicolumn{1}{c}{$\Xi^- \times10^5$} & $\Omega^- \times10^4$  \\
(\textit{A} GeV) & (Million) \\ 
\hline
10   & 4.6 & 7.37 & 1.74 & 6.70 & 9.76 & 7.68 & 2.73  \\ 
20  & 3.1  & 7.66 & 2.72 & 6.40 & 17.81 & 11.80 & 6.30  \\
30  & 4.9 & 14.28 & 6.55 & 15.15 & 42.36 & 24.20 & 20.07  \\
40  & 1.1 & 3.47 & 1.82 & 3.51 & 11.02 & 6.31 & 6.12 \\
\hline
\hline
\end{tabular*}
\end{table*}

\begin{figure*}[tb]
\centering
\includegraphics*[width=2.0\columnwidth]{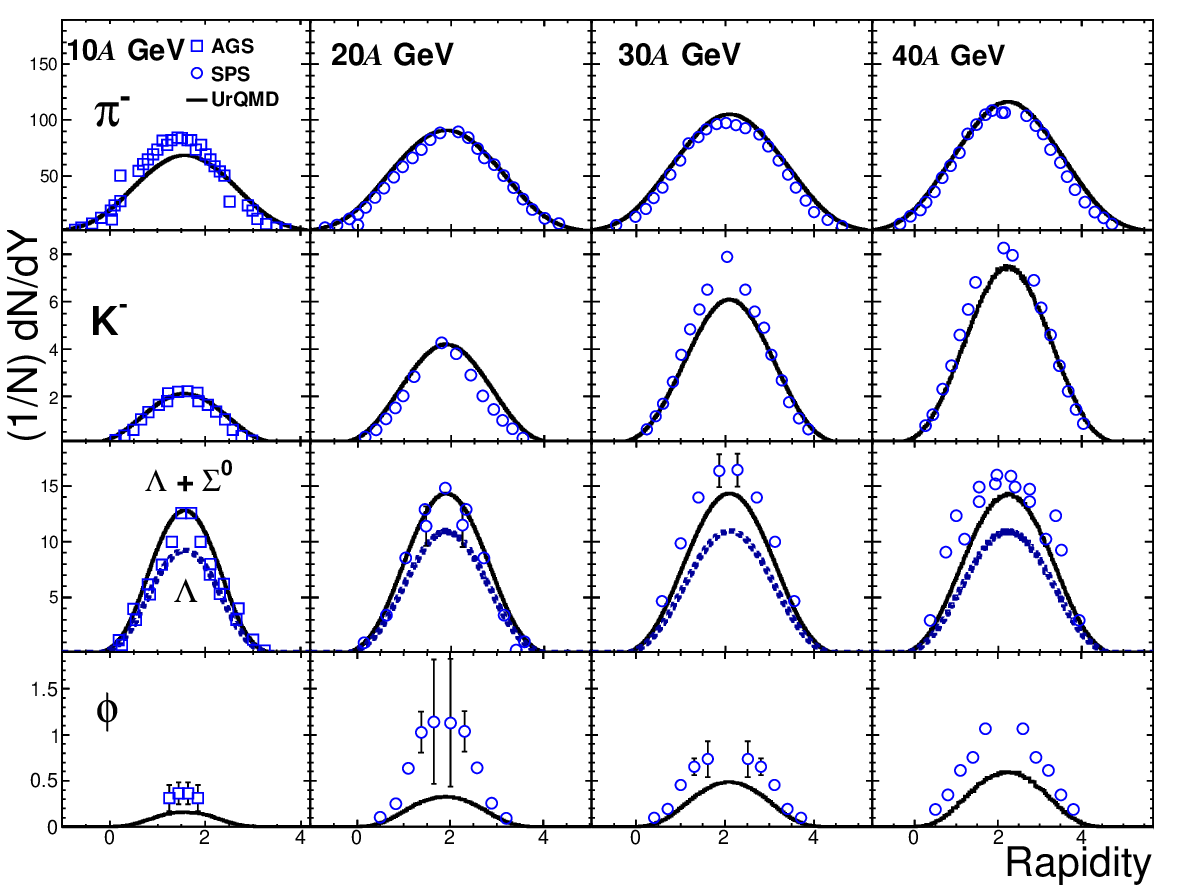}
\caption{(Color online) Rapidity distribution of produced particles $(\pi^{-}, k^{-}, \Lambda , \phi)$ for four different beam energies $E_{LAB}$= 10, 20, 30 and 40\textit{A} GeV is compared with the existing SPS (NA49) \cite {data1, data2, data3, data4, data5} (7.2$\%$ central) and AGS (E877, E802, E896, E917) \cite{AGS1, AGS2, AGS3, AGS4, AGS5, AGS6, AGS7} data ($5\%$ central). It is to be noted that the NA49 data (20-40\textit{A} GeV) are for the $Pb+Pb$ system and the AGS data (10\textit{A} GeV) are for the $Au+Au$ system, while the UrQMD calculation is done for the $Au+Au$ system with impact parameter 0-3 fm (5$\%$) and 0-3.5 fm (7.2$\%$).}
\label{rapidity}
\end{figure*}

\section{Results on width of rapidity distribution}
In a heavy ion collision, the rapidity $(y)$ or pseudorapidity ($\eta$) distribution is found to be quite informative of the particle production mechanism \cite{width, rapidity}. In Fig.\ref{rapidity}, the rapidity distributions of a few hadrons such as  $\pi{^-}$, $k{^-}$, $\phi$, and $\Lambda$ produced in  UrQMD generated $Au+Au$ collisions at 10, 20, 30 and 40\textit{A} GeV are plotted and compared with the existing Super Proton Synchrotron (SPS) \cite {data1, data2, data3, data4, data5} and Alternating Gradient Synchrotron (AGS) (E802, E877, E896, E917) \cite{AGS1, AGS2, AGS3, AGS4, AGS5, AGS6, AGS7} data at the same energies. It is seen from this figure that, with UrQMD generated events, the rapidity distribution of ($\Lambda$ + $\Sigma^{0}$) and not $\Lambda$ alone gives a better agreement with the experimental data on $\Lambda$. This is because of the fact that $\Lambda$, resulting from the decay of $\Sigma^{0}$, cannot be separated from the directly produced ones via a secondary vertex measurement \cite{lambda_sigma} and hence the experimentally measured value of $\Lambda$ is an overestimation of the number of $\Lambda$ actually produced in the collision.

From this figure it is also seen that lighter mesons like $\pi^-$ and $K^-$ show a fair agreement between experimental data \cite{data1, data2, data3, data4, data5, AGS1, AGS2, AGS3, AGS4, AGS5, AGS6, AGS7} and UrQMD prediction at all energies. However, for heavier meson like $\phi$, as the energy of collision increases, there is a considerable disagreement between the observed and UrQMD predicted values. This has been attributed to the fact that \cite{data3}, at low energies, $E_{LAB}$ $\leq$10\textit{A} GeV, the $\phi$ production mechanism is predominantly via hadronic channels; at higher energies there might be significant contribution from nonhadronic processes as well. In UrQMD, with regard to the production of $\phi$ meson, the hadronic processes mostly refer to KK coalescence whereas other nonhadronic contributions mostly come from string excitation and fragmentation.

The rapidity spectra of all the studied hadrons from UrQMD generated and experimental data \cite{AGS1, AGS2, AGS3, AGS4, AGS5, AGS6, AGS7} (AGS data) are parametrized by a sum of two Gaussian functions displaced symmetrically around midrapidity by a shift $a$: 

\begin{equation}
\frac{dn}{dy} \propto  \left[ e^{-\frac{(y-a)^2}{2\sigma^2}} + e^{-\frac{(y+a)^2}{2\sigma^2}} \right] 
\end{equation}

The width of the distribution is characterized by root mean square RMS $ = \sqrt{\sigma^2+a^2}$  and the resulting values of RMS are listed in Table \ref{fit}.\\

\begin{table*}[tb]
\setlength\extrarowheight{3pt}
\caption{The parameter RMS values resulting from the double Gaussian fits of the rapidity spectra of all the studied hadrons of UrQMD generated and experimental data (within parentheses). The RMS values of NA49 data are taken from refs. \cite{data5, data3, data4}. }
\label{fit}
\begin{tabular*}{\textwidth}{@{\extracolsep{\fill}}lcccc@{}}
\hline
\hline
RMS  & \multicolumn{1}{c}{10\textit{A} GeV} & \multicolumn{1}{c}{20\textit{A} GeV} & \multicolumn{1}{c}{30\textit{A} GeV} & \multicolumn{1}{c}{40\textit{A} GeV}  \\
\hline 
$\pi^-$ & 0.959 $\pm$ 0.00004 & 1.050 $\pm$ 0.00005 & 1.111 $\pm$ 0.00003 & 1.149 $\pm$ 0.00007\\
			     & (0.844$\pm$ 0.08) & (0.991 $\pm$ 0.01) \cite{data5} & (1.068 $\pm$ 0.01) \cite{data5} & (1.123 $\pm$ 0.01) \cite{data5}\\ 
			     	   
$K^-$ & 0.716 $\pm$ 0.0002  & 0.803 $\pm$ 0.0001 & 0.854 $\pm$ 0.0001 & 0.887 $\pm$ 0.0002\\
			   & (0.628 $\pm$ 0.02)  & (0.727 $\pm$ 0.034) \cite{data5} & (0.798 $\pm$ 0.009) \cite{data5} & (0.852 $\pm$ 0.069) \cite{data5}\\
				
$\phi$ & 0.560 $\pm$ 0.0006 & 0.664 $\pm$ 0.0005 & 0.724 $\pm$ 0.0003 & 0.761 $\pm$ 0.0006\\
				& (0.511 $\pm$ 0.301) & (0.582 $\pm$ 0.031) \cite{data3} & (0.769$\pm$ 0.030) \cite{data3} & (0.852$\pm$0.015) \cite{data3}\\
				
$\Lambda$ & 0.630 $\pm$ 0.00007 & 0.741 $\pm$ 0.0003 & 0.814 $\pm$ 0.0006 & 0.864 $\pm$ 0.0006\\
				   & (0.648 $\pm$ 0.05) & (0.70 $\pm$ 0.01) \cite{data4} & (0.89$\pm$0.02) \cite{data4} &  (1.11$\pm$0.08) \cite{data4}
				   \\
				
$\Xi^-$ & 0.519 $\pm$ 0.0008 & 0.605 $\pm$ 0.0008  & 0.661 $\pm$ 0.0002  & 0.705 $\pm$ 0.0063\\
					&	  		& (0.64 $\pm$ 0.08) \cite{data4} & (0.73 $\pm$ 0.14) \cite{data4} & (0.94 $\pm$ 0.13) \cite{data4}\\
				
$\Omega^-$ & 0.450 $\pm$ 0.0015 & 0.536 $\pm$ 0.0017 & 0.590 $\pm$ 0.0034 & 0.619 $\pm$ 0.004\\
					&                 &                &                & (0.596 $\pm$ 0.09) \cite{data5} \\
\hline
\hline
\end{tabular*}
\end{table*}

\begin{figure*}[tb]
\centering
\includegraphics*[width=2.1\columnwidth]{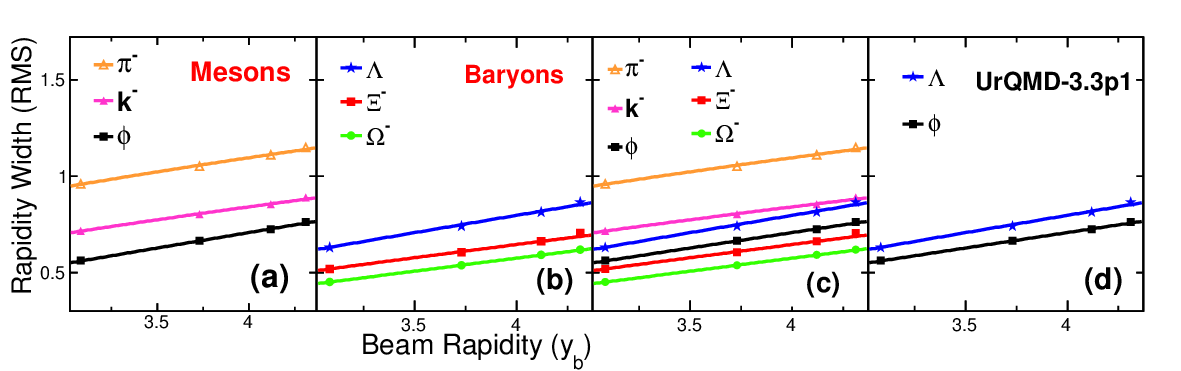}
\caption{(Color online) Variation of width of the rapidity distribution with UrQMD generated data of (a) mesons, (b) baryons, and (c) all studied hadrons as a function of beam rapidity in the laboratory system. In the panel (d) the same has been plotted separately for $\Lambda$ and $\phi$. The solid lines correspond to linear fits. The errors are small and are within the symbol size.}

\label{width}
\end{figure*}

\begin{figure*}[tb]
\centering
\includegraphics*[width=1.6\columnwidth]{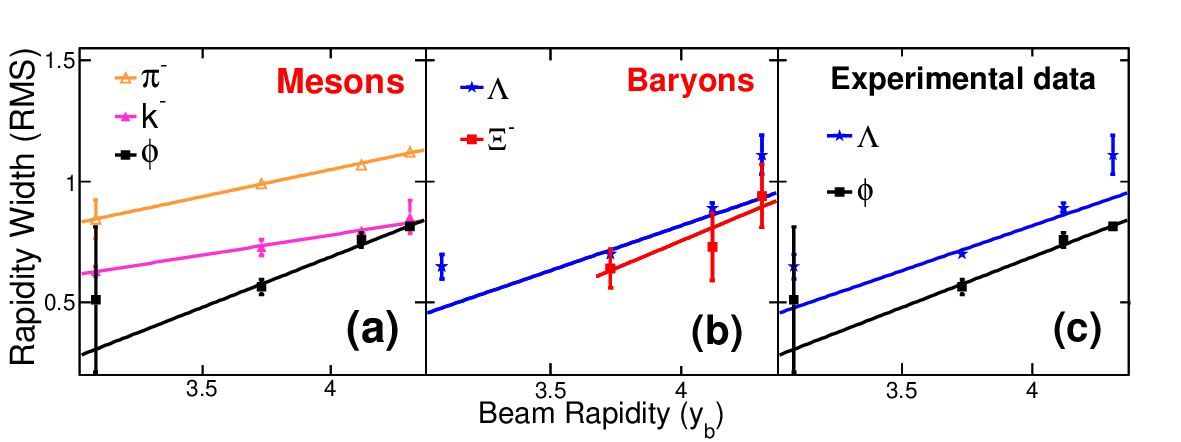}
\caption{(Color online) Variation of width of the rapidity distribution (calculated from data \cite{data1, data2, data3, data4, data5, AGS1, AGS2, AGS3, AGS4, AGS5, AGS6, AGS7}) of (a) mesons and (b) baryons  as a function of beam rapidity in the laboratory system. (c) The same has been plotted separately for $\Lambda$ and $\phi$. The solid lines correspond to linear fits. The error bars shown here correspond to the statistical error. The large error at 10\textit{A} GeV in $\phi$ data is due to large experimental error.}
\label{width_data}
\end{figure*}

\begin{figure}[tb]
\centering
\includegraphics*[width=80mm]{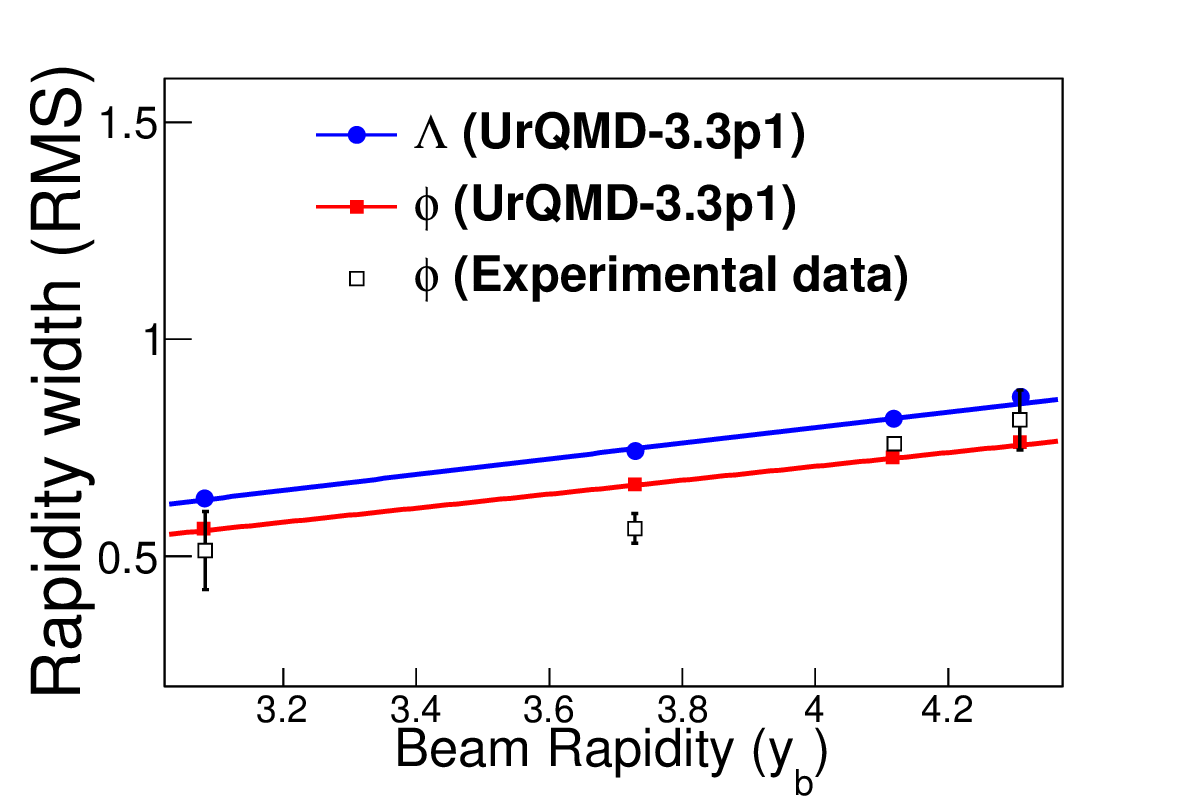}
\caption{(color online) The width of the rapidity distribution as a function of beam rapidity for $\phi$ from UrQMD,  experimental data (AGS+SPS), and $\Lambda$(UrQMD). }
\label{phi_comparisn}
\end{figure}

Figure \ref{width}  represents the width of the rapidity distributions as a function of beam rapidity  for studied mesons and baryons for UrQMD generated central $Au+Au$ collision. As expected from the kinematical point of view, the lighter particles are found to have larger rapidity width. From this plot, a scaling behavior of the width of the rapidity distribution with beam rapidity is readily evident. Friese \textit{et al.} \cite{phi_NA49_2} have reported similar results for Pb+Pb collision with NA49 data. It is interesting to note from Figs. \ref{width}(a) and \ref{width}(b) that, although  mesons and baryons separately follow mass ordering, such mass ordering is violated if the studied hadrons are taken together [Fig. \ref{width}(c)]. Figure \ref{width_data} contains the same plots as Fig.\ref{width}, but with the experimental data (of refs. \cite{data1, data2, data3, data4, data5, AGS1, AGS2, AGS3, AGS4, AGS5, AGS6, AGS7}) of Fig. \ref{rapidity} of this article. It can be readily seen from Fig.\ref{width_data} that for the experimental data also the mass ordering exists only when mesons and baryons are taken separately. Mass ordering gets disturbed if these hadrons are taken together.
To see the possible effect of underestimation of the yield of UrQMD generated $\phi$ mesons on the mass ordering of the width of the rapidity distribution, the widths of the rapidity distributions of the $\phi$ meson, from both UrQMD and experimental data, are plotted in Fig. \ref{phi_comparisn} and compared with the result obtained for $\Lambda$ with UrQMD generated data. Clearly the rapidity width of $\Lambda$ is larger than the rapidity width of UrQMD and experimental $\phi$. Such an observation implies that, even though the experimental values of the rapidity distribution of $\phi$ differ from  UrQMD prediction, such differences do not have any significant effect on our result for the mass ordering violation of the width of the rapidity distribution with beam rapidity for the studied hadrons.

\begin{figure*}[tb]
\centering
\includegraphics*[width=1.55\columnwidth]{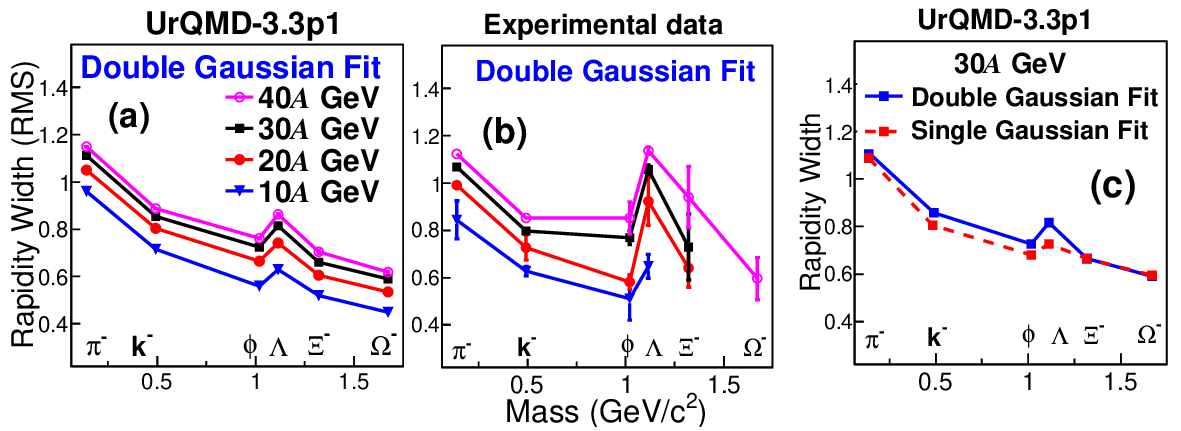} 
\caption{(Color online) Variation of rapidity width as a function of mass of the produced particles in central $Au+Au$ collision at FAIR energies for UrQMD and experimental data. The left and middle panels correspond to double Gaussian parametrization while the right panel corresponds to a comparison of single and double Gaussian fits for 30\textit{A} GeV. The errors seen in the figure correspond to the statistical error.}

\label{width_vs_msaa}
\end{figure*}

In order to have a clearer picture of the mass ordering violation, in Fig. \ref{width_vs_msaa}, the width of the rapidity distribution at a particular energy is plotted as a function of the mass of the produced particles. It is readily seen from this figure that at each energy, for UrQMD generated data, the width of the rapidity distribution follows a separate mass ordering for the studied mesons and baryons.  A similar behavior could be seen with the experimental data as well.
To show that the observed violation of mass ordering is independent of the choice of a particular type of fitting of rapidity spectra, in Fig. \ref{width_vs_msaa}(c) the widths of the rapidity distributions for single and double Gaussian fits are plotted as a function of mass of the produced particle. It is readily evident from this figure that the type of fitting of rapidity spectra has little effect on our observation of separate mass scaling for the studied mesons and baryons.

\begin{figure*}[tb]
\centering
\includegraphics*[width=1.55\columnwidth]{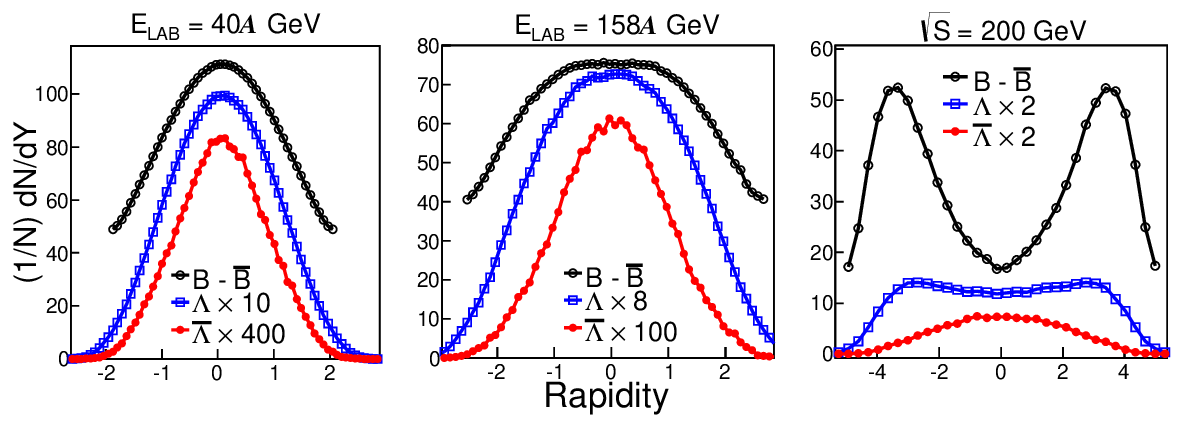} 
\caption{(Color online) Comparison of rapidity distribution of net-baryon number with that of $\Lambda$ and $\bar{\Lambda}$ for $Au+Au$ at $E_{LAB}$ = 40, 158\textit{A} GeV and $\sqrt{s}$ = 200 GeV using UrQMD-3.3p1.}

\label{net_baryon}
\end{figure*}

The observed violation of mass ordering may be attributed to the fact that, at FAIR and SPS energies, due to ``semitransparency", the net baryon density at midrapidity is expected to be intermediate between the SIS18/AGS and RHIC/LHC situations. That is, the net-baryon density is neither Gaussian peaked at mid-rapidity nor vanishingly small. The rapidity distribution of a particle, the production of which is sensitive to net baryon density, will tend to follow the distribution of net-baryon density; i.e. it would be broader than that of a particle (mesons having no leading quark/antibaryons) the production of which does not depend on net-baryon density. The effect would be greater for the baryons that contains more light quarks. Earlier, Bleicher \cite{bleicher} and then Steinheimer and Bleicher \cite{stein} reported the dependence of the excitation function of the width of the rapidity distribution on the initial up- and down-quark content of hadrons. This coupling to net-baryon density comes on top of the kinematic behavior which makes the rapidity distribution narrower for heavy particles. In the case of $\Xi^{-}$ and $\phi$, the two effects seem to more or less compensate each other. To show the dependence of rapidity distribution of the particles containing light quarks on net-baryon density, in Fig. \ref{net_baryon}, the rapidity distribution of net-baryon number is compared with rapidity distributions of $\Lambda$ and $\bar{\Lambda}$. A clear dependence of the $\Lambda$ rapidity distribution on net-baryon density is visible in Fig. \ref{net_baryon}.

\begin{figure}[tb]
\centering
\includegraphics*[width=80mm]{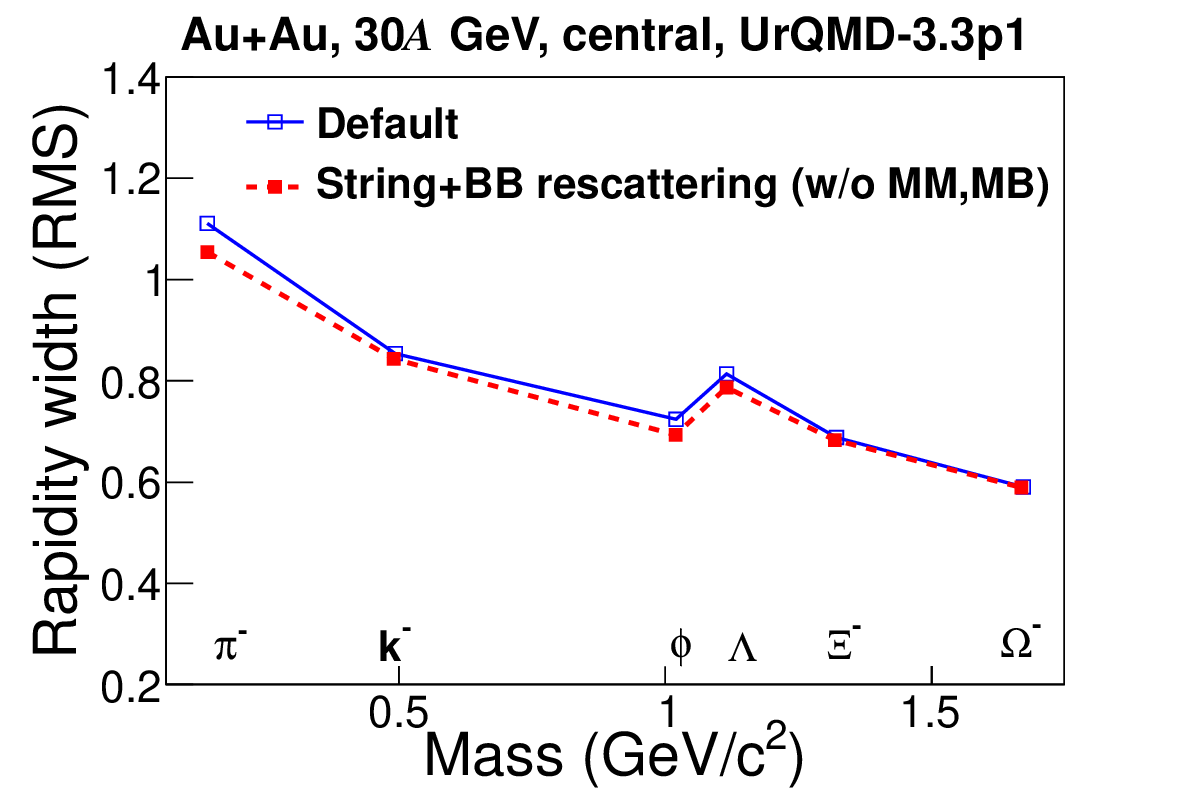} 
\caption{(Color online) Variation of rapidity width as a function of mass of the produced particles for two different scenarios: (a) default (string + BB,MM,MB rescattering), (b) string + BB scattering (without MM,MB rescattering) using UrQMD-3.3p1 at 30\textit{A} GeV.}
\label{width_vs_mass_s}
\end{figure}

\begin{figure}[tb]
\centering
\includegraphics*[width=80mm]{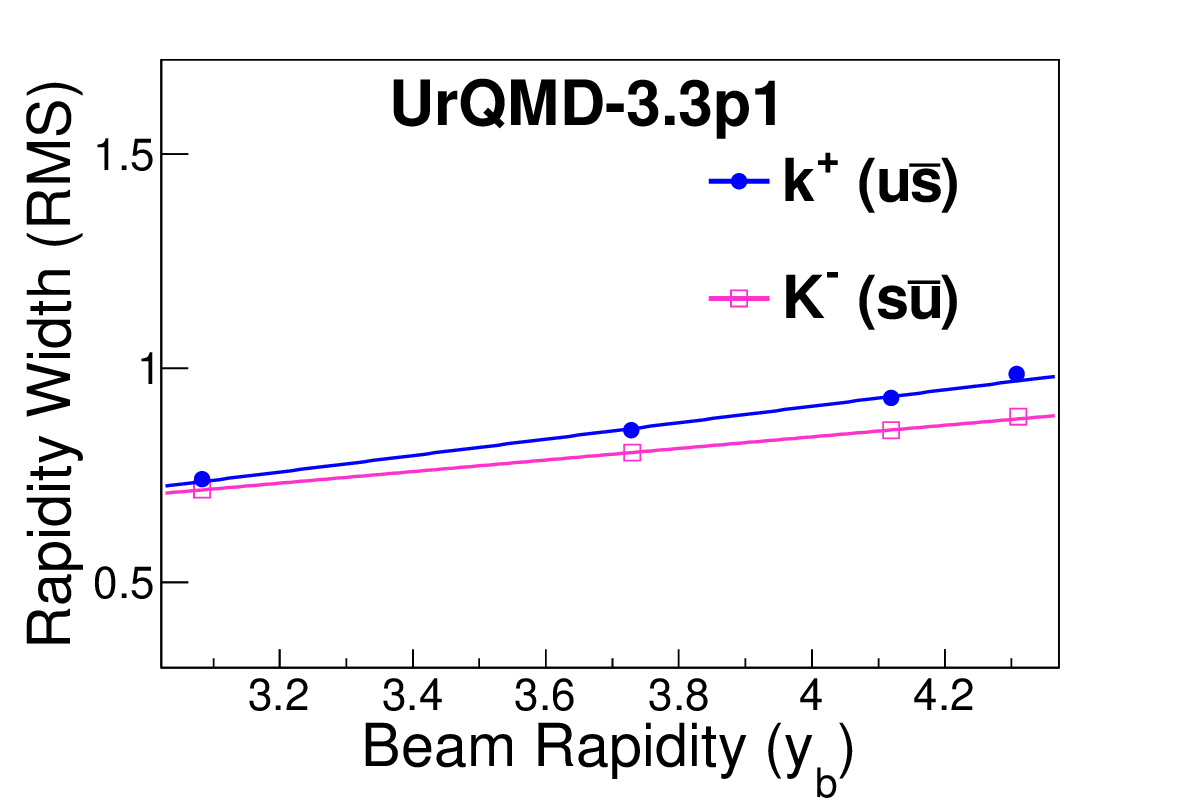} 
\caption{(Color online) Rapidity widths of $K^+$ and $K^-$ as a function of beam rapidity for UrQMD-3.3p1 generated central Au+Au collision.}
\label{kplus}
\end{figure}

In order to ascertain the influence of resonance re-scattering on the width of the rapidity distribution, we regenerated our events at 30\textit{A} GeV by switching off meson-meson (MM) and meson-baryon (MB) rescattering; that is, string excitation and fragmentation and baryon-baryon (BB) rescattering are the main mechanism of particle production. It is to be noted that, in UrQMD, switching off MM and MB rescattering does not turn off all the MM and MB rescattering; it only turns off the processes that occur through an intermediate resonance state like $MM\rightarrow M^{*}\rightarrow MM$ or $MB\rightarrow B^{*}\rightarrow MB$. Even though these are the dominant channels of particle production, elastic and inelastic MM and MB processes are still allowed. This picture of collision implies that the early reactions that are dominated by string excitation and fragmentation still happens in a similar fashion as in the full calculation, but taking away resonance rescattering does indeed shorten the overall reaction time thus decreasing the overall thermalization of the system. Looking at Fig. \ref{width_vs_mass_s} it can be seen that the resonance rescattering seems to have little influence on the width of the rapidity distribution.

\section{Summary}
The present study of the width of the rapidity distribution of mesons and baryons produced in central $Au+Au$ collisions at 10, 20, 30, and 40 \textit{A} GeV using both UrQMD generated and experimental data reveals scaling behavior of the width of the rapidity distribution with beam rapidity. Moreover, the variation of the width of the rapidity distribution with beam rapidity follows a mass ordering separately for mesons and baryons. If the studied hadrons are taken together, the mass ordering is violated. This separate mass ordering is attributed to the fact that, unlike mesons, the width of the baryons' rapidity distribution is being influenced by kinematic consideration as well as the net-baryon density.  In addition, it may be worth mentioning that the rapidity distribution of mesons containing leading quarks, e.g., $K^+$ ($u \overline{s}$), can also be effected by the net-baryon density. This is demonstrated in Fig. \ref{kplus}, where in spite of having the same mass, $K^+$ and $K^-$ show different rapidity widths.

However, considering that the mass ordering of the width of the rapidity distribution could be due to the buildup of some collective flow, the possibility of different strengths of collective flow for mesons and baryons resulting in different mass ordering also cannot be ruled out completely, and needs further investigation.

\vspace*{3mm}

\section*{Acknowledgments}
The authors thankfully acknowledge valuable discussions with Dr. Volker Friese, Dr. S. A. Bass, Dr. Y.P. Viyogi and Dr. Hannah Petersen. This work is supported by Department of Science and Technology, Government. of India through a research project vide project No. SR/MF/PS-012009.

\end{document}